\documentstyle[sprocl,psfig]{article}

\def\be{\begin{equation}}
\def\ee{\end{equation}}
\def\bea{\begin{eqnarray}}
\def\eea{\end{eqnarray}}

\newcommand{\gev}{{\rm GeV}}
\begin{document}
\begin{flushright}
Edinburgh Preprint 96/12\\
Liverpool Preprint LTH-377\\[0.5em]
\end{flushright}
\title{LATTICE STUDY OF THE POMERON}

\author{{\it UKQCD COLLABORATION}\\[0.5em]}
\author{D.S.\ HENTY$^\dagger$, C.\ PARRINELLO$^\ast$, 
D.G.\ RICHARDS$^\dagger$, 
J.I.\ SKULLERUD$^\dagger$\\[0.5em]}
\address{$^\dagger$Dept.\ of Physics \& Astronomy, 
University of Edinburgh, Edinburgh EH9 3JZ, U.K.\\[0.5em]
$^\ast$D.A.M.T.P., University of Liverpool, 
Liverpool L69 3BX, U.K.}


\maketitle\abstracts{We investigate the phenomenology of the Landshoff-Nachtmann two-gluon-exchange
model of the Pomeron using gluon propagators computed in the Landau
gauge within quenched lattice QCD simulations.  As the propagators
have been evaluated entirely from QCD first principles, our results
provide a consistency check of the model.
Finally, we report a proposal to investigate the structure of the
Pomeron-quark vertex, using a colour-singlet Pomeron source
constructed from the gauge fields.}

The Pomeron trajectory was introduced to describe certain features
of hadronic scattering: total hadronic cross sections slowly rising
with energy, their factorisation, the observation that the elastic
amplitude is predominantly imaginary, and the quark-counting rule.
This description has proved remarkably durable, and interest in
pomeron physics has been renewed by recent results from HERA.  In this
talk we will investigate a particular QCD model of the pomeron, the
non-perturbative Landshoff-Nachtmann (LN) two-gluon-exchange (2GE)
model,\cite{ln:87} within a lattice gauge simulation.  We will then
suggest how the Pomeron may be investigated from a lattice gauge
simulation at a more fundamental level.

The 2GE model has proved highly successful at
describing Pomeron exchange.  Crucial to the model is that the gluons
be infra-red finite.  Recently, the
phenomenology has been explored using  gluon propagators
from the solution of the Schwinger-Dyson
equation.\cite{halzen:93,ducati:93} In this talk we will explore the
phenomenology using lattice gluon propagators in the Landau gauge.

The momentum dependence of the gluon propagator we extract from the
lattice simulation.  The ``effective'' coupling of the gluon
propagator to a quark we determine by requiring that the Pomeron-quark
coupling, $\beta_0$, attain its phenomenological value.  We perform
the calculation at several volumes and lattice spacings to control
systematic uncertainties due to the finite volume and non-zero lattice
spacing.\cite{dgr:96}

\begin{figure}[t]
\hspace{1in}\psfig{height=2.7in,figure=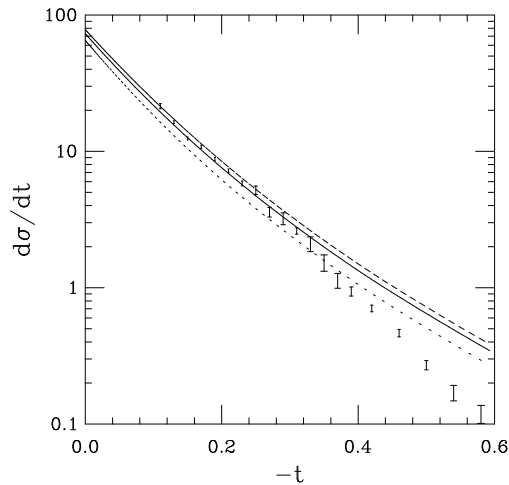}
\caption{Data for the $pp$ elastic cross section at $\protect\sqrt{s} =
53~\gev$ from Ref.~4, together with the lattice
prediction corrected for energy dependence on $16^4$ lattices at
$\beta = 6.2$ (solid) and $\beta = 6.0$ (dots), and on a $24^3
\times 32$ lattice at $\beta=6.0$ (dashes).\label{fig:isr_data}}
\end{figure}

As an example of the calculation, we show in Figure~\ref{fig:isr_data} a
comparison of the model with the $pp$ differential cross section,
using three lattices at different volumes and spacings; both the total
cross section and the slope of the differential cross section are
close to their phenomenological values.
The calculation also exhibits two further features of diffractive
scattering: the quark counting rule, and the weaker coupling of the
Pomeron to mesons composed of heavy quarks.

Encouraged by this application of the LN model, we are tempted to go
further, and explore the nature of the Pomeron-quark
vertex.\cite{vertex} Here we are using a colour-singlet source for the
Pomeron constructed from the gauge fields, and the lattice quark
propagator fixed to the Landau Gauge.  This should provide a valuable
clue to the mechanism for diffractive physics in QCD.

\end{document}